\documentclass[fleqn,twoside]{article}
\usepackage{espcrc2} 
\usepackage{psfig}

\title{Magnetic polarizability of hadrons from lattice QCD
\thanks{presented by L. Zhou at Lattice 2002.}
\thanks{This work is supported in part by U.S. Department of Energy
under grant DE-FG02-95ER40907, and by NSF grant 0070836.
The computing resources at NERSC and JLab have been used.}}

\author{ L. Zhou\address[GW]{Center for Nuclear Studies, 
George Washington University, Washington, DC 20052, USA}, 
F.X. Lee\addressmark[GW]\address[JL]
{Jefferson Lab, 12000 Jefferson Avenue, Newport News, VA 23606, USA},
W. Wilcox\address{Department of Physics, 
Baylor University, Waco, TX 76798, USA},
J. Christensen\address{Department of Physics, 
McMurry University, Abilene, TX 79697, USA}
}

\begin{document}

\begin{abstract}
We extract the magnetic polarizability from the 
quadratic response of a hadron's mass shift in progressively small 
static magnetic fields.
The calculation is done on a 24 $\times$ 12
$\times$ 12 $\times$ 24 lattice at $a$ = 0.17 fm with an 
improved gauge action and the clover quark action. The 
results are compared to those from experiments and models where available. 
\vspace{1pc}
\end{abstract}

\maketitle

Polarizabilities are important fundamental properties of particles. They 
determine dynamical response of a bound system to external perturbations, 
and provide valuable insight into internal strong interaction structure. 
We discuss a lattice calculation for static magnetic 
polarizability $\beta$ in the quenched approximation. This work is an
extension of a lattice calculation for the electric 
polarizability~\cite{walter89}. For an updated calculation of the electric
polarizability, see~\cite{joe02}.

For small external magnetic fields, $\beta$ can be defined via the 
energy (mass) shift of a particle in the fields
\begin{equation}
\Delta m = - \frac{1}{2} \beta {\bf B}^2
\end{equation}
%
We determine the magnetic polarizability by calculating the mass shift 
of hadrons 
\begin{equation}  \label{change}
\Delta m = m(B) - m (0)
\end{equation}
where $m(B)$ is the hadron mass in the presence of an magnetic field $B$. We fit the data
with the polynomial
\begin{equation}  \label{mshift}
\Delta m (B) = c_1 B + \frac{1}{2} c_2 B^2
\end{equation}
We calculate mass shifts both in the field $B$ and its 
inverse $-B$, then average them.
The magnetic polarizabilities are the negative quadratic coefficients
\begin{equation}
\beta = - c_2
\end{equation}

The gauge action is the zero-loop, tadpole-improved 
L\"{u}scher-Weiz action with a=0.17 fm (or 1/a=1159 MeV) set from the string tension.
The tadpole factor is determined by the average plaquette as $u_0=0.877.$
The clover quark action was used to put fermions on the lattice, 
\begin{equation} \label{action}
S_f = S_W - {\kappa \over u_0^4} \sum_x \sum_{\mu < \nu} 
\bigg[ \overline{\psi}(x) ~ i \sigma_{\mu\nu} F_{\mu\nu} \psi(x) \bigg],
\end{equation}
where $S_W$ is the tadpole-improved Wilson fermion action.

In order to place an magnetic field on the lattice, we construct an 
analogy to the continuum case. There, the fermion action is modified 
by the minimal coupling prescription 
$\partial_\mu \rightarrow \partial_\mu + i q A_\mu$, 
where $q$ is the charge of the fermion field and $A_\mu$ is the vector 
potential describing the external field. On the lattice, the prescription
amounts to a phase factor $e^{i a q A_\mu}$ to the link variables~\cite{walter89}.
Choosing $A_2 = B x_1 $ and $A_0 = A_1 = A_3=0$  a constant magnetic field 
$F_{12} = \partial_1 A_2 - \partial_2 A_1 =B_3=B$ can be introduced 
in the $z$-direction. 
We introduce two dimensionless parameters: one is 
\begin{equation} \label{parameter}
\eta = q B a^2,
\end{equation}
where $q= Q e$ with $Q = \pm 1/3, \pm 2/3$. 
The other is the integer lattice length,
\begin{equation}
\rho = x_1/a.
\end{equation}
In terms of these two parameters the phase factor becomes:
\begin{equation}
e^{i a q A_2} = e^{i \eta \rho} \rightarrow (1 + i \eta \rho ), 
\end{equation}
where we have linearized the phase factor to mimic the continuum prescription. 
Therefore, the field strength we use will be chosen to satisfy the linearization requirement.
To summarize, our method to place the external magnetic field on the lattice is to 
multiply each gauge field link variable $U_\mu (x) $ in the 
$z$-direction with a $x$-dependent factor:
\begin{equation} \label{prescription}
U_3 (x) \rightarrow (1 + i \eta \rho ) U_3 (x)
\end{equation}
This factor is applied only to the $S_W$ part of the action, the clover term
is untouched. 
%
%
\begin{figure}
\centerline{\psfig{file=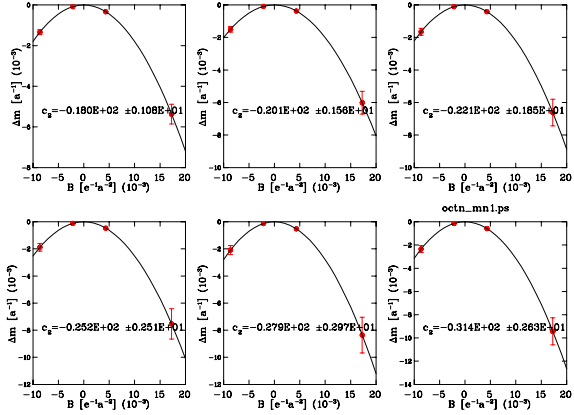,width=7.5cm,angle=90}}
\vspace*{-0.5cm}
\caption{Mass shifts of the neutron as a function of the magnetic field 
$B$ at the six quark masses.}
\label{Fitting_octn}
\vspace*{-0.5cm}
\end{figure}

Periodic boundary conditions were imposed on the fermion fields 
in $y$ and $z$ directions.  In the $x$ and $t$ directions, 
we have employed fixed (or Dirichlet) boundary conditions, 
i.e. the fermion fields are not allowed to propagate 
across the edges of $x$ and $t$.
To minimize the non-periodic boundary effects, we use a lattice size of 
$24 \times 12 \times 12 \times 24$ 
with the fermion source location $(x,y,z,t)=(12,1,1,3)$.
We analyzed 100 configurations. 

Fermion propagators $M^{-1}$ were constructed at six $\kappa$ values, 
$\kappa$=0.1219, 0.1214, 0.1209, 0.1201, 0.1194, 0.1182 
corresponding to $m_\pi$=450 to 820 MeV.
The critical value $\kappa_c=0.1232(1)$.
We use six different values of the parameter 
$\eta$ = 0.0, $+0.00036$, $-0.00072$, $+0.00144$, $-0.00288$, $+0.00576$.
The $\eta$ values in this sequence are related by a factor of $-2$. 
Thus we are able to study the response of a 
hadron composed of both up and down (strange) quarks to four different nonzero
magnetic fields. 

\begin{figure}
\centerline{\psfig{file=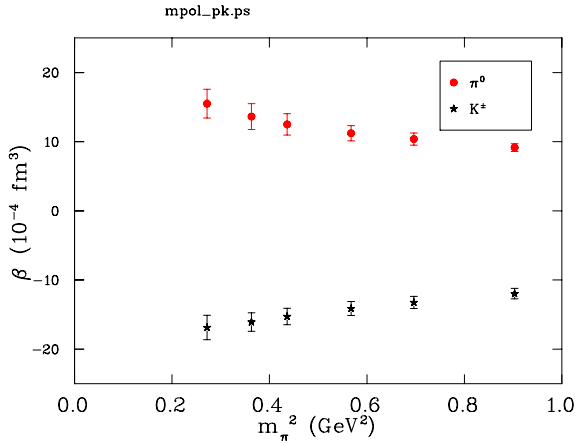,width=7.5cm,angle=90}}
\vspace*{-0.5cm}
\caption{Magnetic polarizability for pion and kaon.}
\label{mpol_pk}
\vspace*{-0.5cm}
\end{figure}

Typical mass shifts are displayed in
Fig.~\ref{Fitting_octn} in the case of neutron.
There is good parabolic behavior going through the origin, 
an indication that contamination from the linear term has been 
effectively eliminated by averaging results from $B$ and $-B$.
%
\begin{figure}
\centerline{\psfig{file=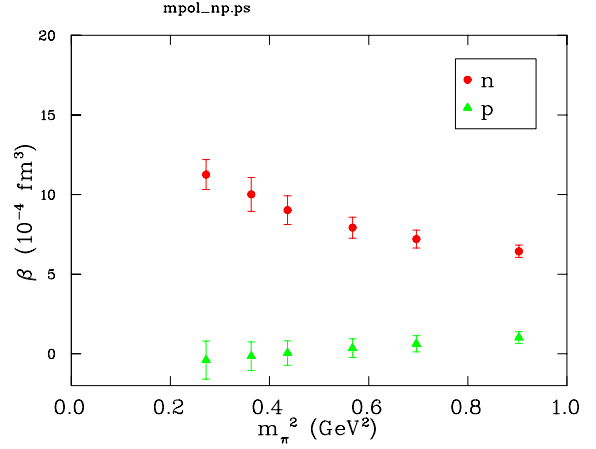,width=7.5cm,angle=90}}
\vspace*{-0.5cm}
\caption{Magnetic polarizability of the neutron and proton as a function 
of $m_\pi^2$ in physical units.}
\label{mpol_np}
\vspace*{-0.5cm}
\end{figure}
The factor 
\begin{equation}
e^2 a^3 = 0.358 \times 10^{-43} {\rm cm}^3
\end{equation}
was used to convert the lattice numbers into 
physical polarizability in units of $10^{-4} {\rm fm}^3$. 
The curves in Fig.~\ref{mpol_np} and Fig.~\ref{mpol_pk} represent 
our preliminary results for neutron, proton, $\pi^0$, and K$^\pm$. 
They are smooth functions of the quark mass.
We did not attempt an extrapolation in the quark mass.

In Table~\ref{nppk}, we show some experimental and
theoretical values for the magnetic polarizabilities $\beta$ of
neutron, proton, pion and kaon.
%
The $\beta$ for neutron is not well-determined.
A recent experimental value~\cite{kossert02} is 2.7 with a big
error bar, while an earlier experimental data analysis \cite{levchuk01}
has it at 11. The range from models and experiments
is from 1.22 to 14. Our lattice simulation result is at the higher end of this range. 
For the proton, the range is 1.6 to 4.4, while our lattice result lies in the range.
For $\pi^0$, there is scant data.
The number from Chiral Perturbation Theory calculation to 
${\mathcal{O}} (p^6)$ is 0.5 to 1.7.
Our result is quite high, about 20.
For $K^+$, there is one theoretical value $-6.2$ from the NJL model. 
Our result is around -20.

In summary, we have computed the magnetic polarizability of hadrons 
for the first time on the lattice using the external field method.
In additon to the particles described here, we have computed $\beta$ for other mesons, 
the other members of the octet baryon, and the entire decuplet baryon family. 
We plan to present the final results elsewhere.
Also under way is the extraction of magnetic moments from the same data set.

\begin{center}
\begin{table}  
\caption{A compilation of magnetic polarizabilitiy for selected hadrons
from experiments or models. The numbers are in units of $10^{-4}$ fm$^3$. \label{nppk}}
\begin{tabular}{cccc}
\hline
Ref.           &  Particle  & $ \beta$                & Approach  \\
\hline \hline
\cite{kossert02}    & n & 2.7 $\mp 1.8^{+0.6}_{-1.1}\mp 1.1$ & Theo.  \\
\cite{liebl95}      & n & $6.7^{+1.3}_{-0.7}$         & Theo.  \\
\cite{bernard93}    & n & 7.8                         & Theo.  \\
\cite{hornidge00}     & n & 10.3 $\mp$ 2.0              & Exp.   \\
\cite{levchuk01}    & n & 11 $\pm$ 3                  & Theo.  \\
\cite{federspiel91} & p & 3.3  $\mp$ 2.2 $\mp$ 1.3    & Exp.   \\
\cite{macgibbon95}  & p & 2.1  $\mp$ 0.8 $\mp$ 0.5    & Exp.   \\
\cite{bernard93}    & p & 3.5                         & Theo.  \\
\cite{hallin93}     & p & 4.4 $\pm$ 0.4 $\pm$ 1.1     & Exp.   \\
\cite{zieger92}     & p & ${3.58 ^{+1.19}_{-1.25}} ^{+1.03}_{-1.07} $  & Exp.   \\
\cite{bellucci94}   & $\pi^0$ &  0.50                 & Theo.  \\
\cite{bellucci94}   & $\pi^0$ &  1.50 $\pm$ 0.20       & Theo.  \\
\cite{ebert97}      & $K^+$ &  -6.2 (-2.9)            & Theo. \\
\hline
\end{tabular}
\end{table}
\end{center}
\vspace*{-1cm}


\end{document}